\documentclass[twocolumn]{pasj00}
\SetRunningHead{TSUMURA et al.}{Extragalactic Background Light Spectrum with AKARI IRC}
\Received{2013 March 27}
\Accepted{2013 July 25}
\Published{}         

\newcommand{\nw}{nWm$^{-2}$sr$^{-1}$}

\begin{document}
\title{Low-Resolution Spectrum of the Extragalactic Background Light with AKARI InfraRed Camera} 
\author{Kohji \textsc{Tsumura}\altaffilmark{1}, Toshio \textsc{Matsumoto}\altaffilmark{1,2}, Shuji \textsc{Matsuura}\altaffilmark{1}, Itsuki \textsc{Sakon}\altaffilmark{3}, and  Takehiko \textsc{Wada}\altaffilmark{1}}
\altaffiltext{1}{Department of Space Astronomy and Astrophysics, Institute of Space and Astronautical Science, Japan Aerospace Exploration Agency, 3-1-1 Yoshinodai, Chuo-ku, Sagamihara, Kanagawa 252-5210}
\altaffiltext{2}{Institute of Astronomy and Astrophysics, Academia Sinica, No.1, Roosevelt Rd, Sec. 4, Taipei 10617, Taiwan, R.O.C.}
\altaffiltext{3}{Department of Astronomy, Graduate School of Science, The University of Tokyo, Hongo 7-3-1, Bunkyo-ku, Tokyo 113-0033}
\email{tsumura@ir.isas.jaxa.jp}
\maketitle
 
\begin{abstract}
The Extragalactic Background Light (EBL) as an integrated light from outside of our Galaxy includes information of the early universe and the Dark Ages. 
We analyzed the spectral data of the astrophysical diffuse emission obtained with the low-resolution spectroscopy mode on the AKARI Infra-Red Camera (IRC) in 1.8-5.3 $\mu$m wavelength region. 
Although the previous EBL observation in this wavelength region is restricted to the observations by DIRBE and IRTS, 
this study adds a new independent result with negligible contamination of Galactic stars owing to higher sensitivity for point sources. 
Other two major foreground components, the zodiacal light (ZL) and the diffuse Galactic light (DGL), were subtracted by taking correlations with ZL brightness estimated by the DIRBE ZL model and with the 100 $\mu$m dust thermal emission, respectively.  
The isotropic emission was obtained as EBL, which shows significant excess over integrated light of galaxies at $<$4 $\mu$m.
The obtained EBL is consistent with the previous measurements by IRTS and DIRBE. 
\end{abstract}
 
\section{Introduction}
The Extragalactic Background Light (EBL) is the integrated light of all light sources outside of our Galaxy. 
At near-infrared (NIR) wavelengths, the dominant physical process for the generation of photon is thought to be nucleosynthesis in stars. 
The EBL contains the accumulated history from first stars of universe to stars at the present days. 
Galaxy counts provide necessarily lower limits to the EBL, although unresolved or faint emission sources may also contribute to EBL. 
One of possible and important faint source is the first stars of the universe which reionized the universe. 
Since individual detection of the first stars is extremely difficult even with JWST, 
detection of excess background over integrated light of galaxies has been searched for. 
Separation of the EBL from foreground emission is very difficult due to its diffuse, extended nature. 
The largest uncertainty comes from the removal of the dominant foreground, the zodiacal light (ZL), 
which is the scattered sunlight by interplanetary dust (IPD) at $<$3.5 $\mu$m and thermal emission from the same IPD at $>$3.5 $\mu$m. 

In order to determine the total EBL brightness, absolute measurement of sky brightness from space is inevitable to avoid strong airglow emission. 
The Diffuse Infrared Background Explorer (DIRBE) on the Cosmic Background Explorer ({\sc COBE})  \citep{Hauser98, Cam01}  
and the Near-Infrared Spectrometer (NIRS) on the Infrared Telescope in Space ({\sc IRTS}) \citep{Matsumoto05, Matsumoto2013} indicate 
that the total EBL brightness at NIR after subtraction of ZL and other foregrounds 
significantly exceeds the brightness determined from deep galaxy number counts \citep{Dominguez2011}. 
However, EBL derived from absolute measurements depends critically on the choice of ZL models \citep{Kelsall98, Wright98}. 
The EBL in optical bands has been measured by  \citet{Bernstein07}, \citet{Matsuoka11} and \citet{Mattila11}, however, their results are controversial. 
Beside uncertainty of ZL model, it has been claimed that observation of TeV-$\gamma$ blazar favors low level EBL at NIR \citep{Dwek05, Aha06, Aha07, Mazin07, Raue09}. 
Furthermore, it has been pointed out that vast formation rate during the first star formation era is needed, if we attribute the origin of this excess emission to first stars \citep{Madau05}.

In order to confirm the origin of the excess emission, new space observations have been desired. 
Unfortunately, {\it Spitzer} was unable to perform absolute measurement of the surface brightness of the sky because of the lack of the cold shutter  \citep{Fazio2004}. 
Here, we present the new result of EBL observation with AKARI InfraRed Camera (IRC). 
One advantage of AKARI/IRC observation is its detection limit for point sources ($m_K \sim 19$),
which is much deeper than COBE/DIRBE and IRTS/NIRS owing to the large aperture telescope. 
This makes contribution of Galactic stars to the background radiation almost negligible and subtraction of the foreground emission more reliable.

This paper is organized as follows. 
In Section \ref{sec_reduction}, we describe the data reduction. 
In Section \ref{sec_foregrounds}, we describe our method for subtraction of foreground emissions, and the resultant EBL spectrum is shown in Section \ref{sec_result}. 
Discussions and implications of our data are given in Section \ref{sec_discussion}.
There are two companion papers describing the spectrum of the infrared diffuse foregrounds; 
data reduction method and ZL is described in \citet{Tsumura2013a} (hereafter Paper I) and the Diffuse Galactic Light (DGL) in \citet{Tsumura2013b} (hereafter Paper II).
Results in Paper I and II are used in this paper (Paper III) for the foreground subtraction.

\section{Data Selection and Reduction} \label{sec_reduction}
AKARI is the first Japanese infrared astronomical satellite launched on February 2006, equipped with a cryogenically cooled telescope of 68.5 cm aperture diameter \citep{Murakami07}.
IRC is one of two astronomical instruments of AKARI, and it covers 1.8-5.3 $\mu$m wavelength region with a 512$\times $412 InSb detector array in the NIR channel\footnote{IRC has two other channels covering 5.8-14.1 $\mu$m in the MIR-S channel and 12.4-26.5 $\mu$m in the MIR-L channel.} \citep{Onaka07}.
It provides low-resolution ($\lambda /\Delta \lambda \sim 20$) slit spectroscopy for the diffuse radiation by a prism\footnote{High-resolution spectroscopy ($\lambda /\Delta \lambda \sim 120$) with a grism is also available.} \citep{Ohyama2007}.

See Paper I for the details of the data selection and reduction.
According to our criteria of data selection for the diffuse background analysis (Paper I), 
total number of 278 diffuse spectra toward randomly distributed sky directions in wide ranges of ecliptic and Galactic coordinates were selected.
Filter wheel of the IRC instrument has dark position to measure the dark current, 
while the cold shutter of {\it Spitzer} has not been operated in the orbit \citep{Fazio2004}. 
Uncertainty due to dark current subtraction is estimated to be $<3$ \nw\ at 2 $\mu$m \citep{TsumuraWada2011}.

Point sources brighter than $m_K(\textrm{Vega}) = 19$ were detected on the slit and masked for deriving the diffuse spectrum.
It was confirmed that the brightness due to unresolved Galactic stars under this detection limit is negligible ($<$0.5 \% of the sky brightness at 2.2 $\mu$m) by a Milky Way star count model, TRILEGAL \citep{Girardi2005}.
This is a great advantage to the previous measurements by DIRBE and IRTS, because the integrated light from unresolved Galactic stars for those measurements are not negligible.
For example in IRTS case, the limiting magnitude for the removing point sources was 10.45 mag at 2.24 $\mu$m, and the contribution of the integrated light from unresolved Galactic stars under this limiting magnitude was $\sim$10\% of the observed sky brightness at high ecliptic latitude \citep{Matsumoto05}, 
which was subtracted by the SKY model for the Galactic point sources \citep{Cohen1994}.
The contamination from the unresolved Galactic stars for the DIRBE measurement was $\sim$25\% at 2.2 $\mu$m \citep{Hauser98}, 
which is greater than that of IRTS owing to the worse detection limit of DIRBE.

\begin{figure}
  \begin{center}
    \FigureFile(80mm,50mm){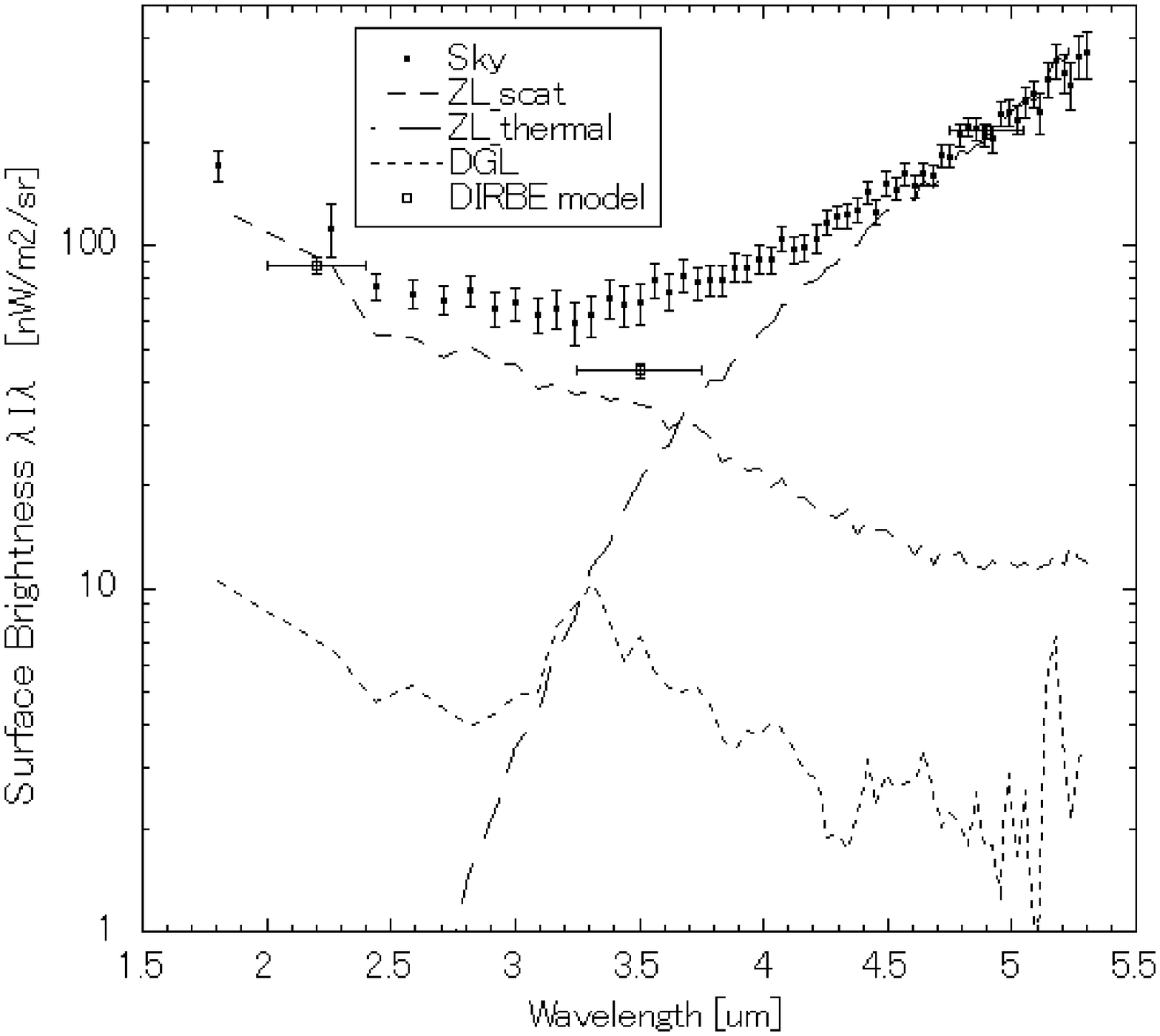}
  \end{center}
  \caption{An example of the sky spectrum at NEP with foreground components.
  Solid circles indicates the sky spectrum in our dataset obtained by AKARI IRC.
  Broken line and dashed line indicate the ZL spectrum (scattered sunlight component and thermal emission component) scaling to the brightness from the DIRBE ZL model (open squares), and dotted line indicates the DGL spectrum.}
  \label{sky}
\end{figure}

The obtained diffuse sky spectrum includes ZL, DGL, and EBL, i.e.,
\begin{equation} SKY_i (\lambda ) = ZL_i (\lambda ) + DGL_i (\lambda ) +EBL_i(\lambda )  \end{equation}
where $i$ is the data index.   
Cumulative brightness contributed by unresolved galaxies can be estimated by the deep galaxy counts, being $<$4 \nw\ at K band in the case of limiting magnitude of $m_K = 19$ \citep{Keenan10}, which is included in EBL.
Figure \ref{sky} shows an example of sky spectrum at NEP with each foreground components estimated by the methods introduced in Paper I and Paper II.
In the next section, we summarize the methods of foreground subtraction, which are essential to derive EBL.

\section{Subtraction of Foregrounds} \label{sec_foregrounds}
\subsection{Estimation of Diffuse Galactic Light}
The method to estimate DGL spectrum in our dataset is developed in Paper II.
This method can be separated into two stages; the first is to derive the spectral shape of DGL as a template ($DGL_{\textrm{temp}}(\lambda )$) by a correlation to the 100 $\mu$m dust thermal emission $I^{100 \mu m}$ from \citet{Schlegel1998},
and the next is to scale this DGL template spectrum by using the relation between $I^{100 \mu m}$ and 3.3 $\mu$m PAH band emission $E_{3.3}$, i.e.
\begin{equation}  DGL_i(\lambda ) =  E_{3.3}(I_i^{100 \mu m}) \cdot DGL_{\textrm{temp}}(\lambda )  \label{eq_DGL}  \end{equation} 

The key point in the first step is that only DGL has the correlation to the 100 $\mu$m emission from the interstellar dust.
Thus $SKY_i (\lambda ) - ZL_i (\lambda ) $ at each wavelength are correlated to the dust emission at 100 $\mu$m,
and this correlated component can be derived as DGL from the sky spectrum.
Here, $ ZL_i (\lambda ) $ is modeled in Paper I and summarized in the next subsection.
The derived DGL spectrum has a distinct PAH band feature at 3.3 $\mu$m as shown in Figure \ref{sky}.

The 3.3 $\mu$m PAH band feature is distinctive at the bottom of the sky spectrum in our dataset at high Galactic latitude regions ($\mid b \mid <15^{\circ}$),
and the good correlation between the 100 $\mu$m dust thermal emission $I^{100 \mu m}$ and the 3.3 $\mu$m PAH band emission $E_{3.3}$ was confirmed in Paper II.
Assuming the spectral shape of the template DGL spectrum is isotropic at low Galactic latitude regions,
DGL spectrum at each field $DGL_i(\lambda )$ is derived by scaling the DGL template spectrum using the relation $E_{3.3}(I^{100 \mu m})$ between the PAH band intensity and the dust thermal emission.
In the low DGL region at high Galactic latitude ($\mid b \mid >30^{\circ}$) used in this study, 
this relation can be approximated to a linear relation,  $E_{3.3}(I^{100 \mu m}) \propto I^{100 \mu m}$.

\subsection{Estimation of Zodiacal Light}
The method to estimate ZL spectrum in our dataset is developed in Paper I.
Similar to the method of DGL, the method of ZL estimation has also two stages; deriving the template ZL spectrum and scaling it.
The template ZL spectrum is derived by differencing the DGL subtracted spectra at the ecliptic plane (ZL strongest region) and that at NEP (ZL weakest region),
because only ZL depends on ecliptic latitude and isotropic EBL is removed by differencing.
In the wavelength range of this study (1.8-5.3 $\mu$m), ZL includes both scattered sunlight component ($\sim$5800 K) and thermal emission component ($\sim$300 K).
We confirmed in Paper I that the spectral shape of each component does not depend on location while their brightness ratio depends on location.
Therefore, the template spectrum of each component was derived separately ($ZL_{\textrm{temp}}^{\textrm{scat}}(\lambda )$ and $ZL_{\textrm{temp}}^{\textrm{thermal}}(\lambda )$).
The scaling of these template ZL spectra are based on the DIRBE ZL model \citep{Kelsall98}, providing the model ZL intensities at 2.2 $\mu$m and 4.9 $\mu$m ($DIRBE_i^{2.2 \mu m}$ and $DIRBE_i^{4.9 \mu m}$) for each field at any observed time.
\begin{equation}  ZL_i(\lambda ) =  ZL_i^{\textrm{scat}}(\lambda ) + ZL_i^{\textrm{thermal}}(\lambda ) \label{eq_ZL} \end{equation}
\begin{equation}  ZL_i^{\textrm{scat}}(\lambda ) =  DIRBE_i^{2.2 \mu m} \cdot ZL_{\textrm{temp}}^{\textrm{scat}}(\lambda ) \end{equation}
\begin{eqnarray}
ZL_i^{\textrm{thermal}}(\lambda )  =    \nonumber  \\ 
& \hspace{-23mm}  [DIRBE_i^{4.9 \mu m} - ZL_i^{\textrm{scat}}(4.9 \mu m)]  \cdot ZL_{\textrm{temp}}^{\textrm{thermal}}(\lambda )  
\end{eqnarray} 
The estimated ZL brightness at 3.5 $\mu$m by this method is consistent with that from the DIRBE ZL model.
The model uncertainty of the DIRBE ZL model is 6 \nw\ at 2.2 $\mu$m, 2.1 \nw\ at 3.5 $\mu$m, and 5.9 \nw\ at 4.9 $\mu$m \citep{Kelsall98}.

\subsection{Correlation analysis}
\begin{figure*} 
  \begin{center}
    \FigureFile(160mm,100mm){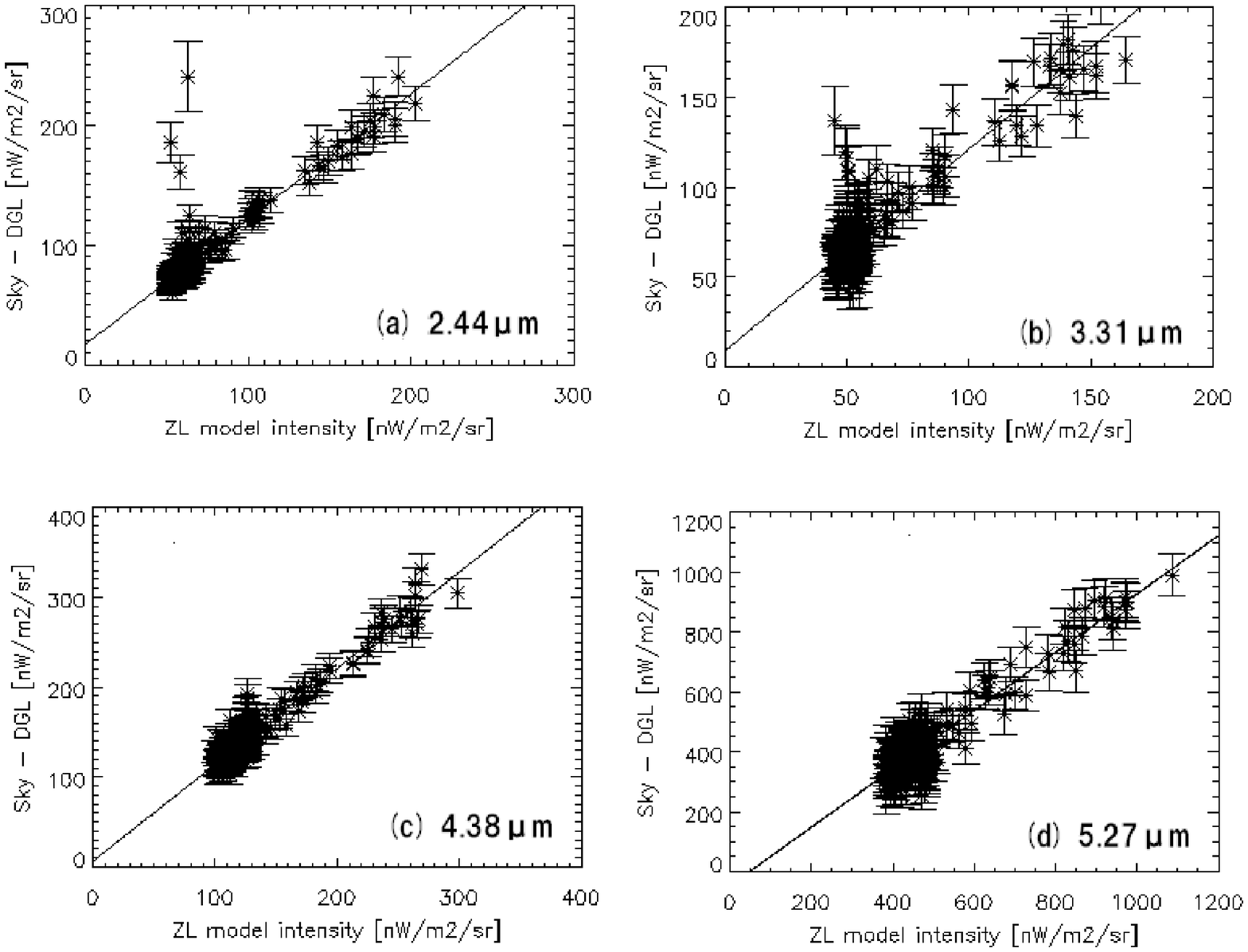}
  \end{center}
  \caption{Examples of correlation between $SKY_i(\lambda ) - DGL_i(\lambda )$ and $ZL_i(\lambda )$ from equation (\ref{eq_ZL}) at (a) 2.44 $\mu$m, (b) 3.31 $\mu$m, (c) 4.38 $\mu$m, and (d) 5,27 $\mu$m.
               The best fit lines are also shown, and error bars in these plots include both statistic random error and the calibration error.}
  \label{correlation}
\end{figure*}

\begin{figure}
  \begin{center}
    \FigureFile(80mm,50mm){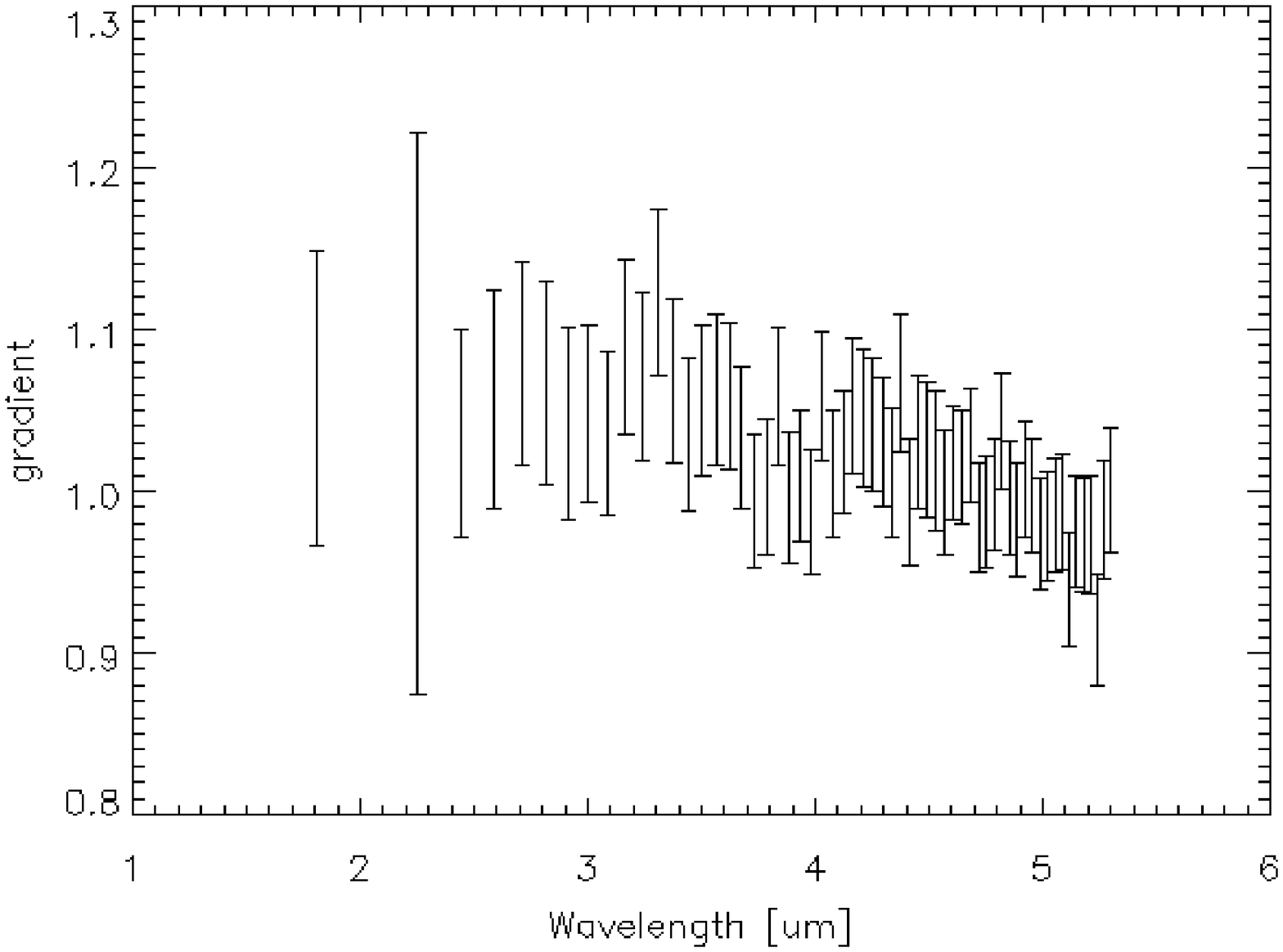}
  \end{center}
  \caption{Gradients of the correlation between $SKY_i(\lambda ) - DGL_i(\lambda )$ and $ZL_i(\lambda )$ shown in Figure \ref{correlation}.
  This is used as a correction factor of ZL, $C(\lambda)$.
  The large error bars especially at short wavelengths are dominated by the calibration error.}
  \label{ZL_correction}
\end{figure}

To derive the isotropic component, we employed a correlation analysis.
First, data at Galactic plane ($\mid b \mid <30^{\circ}$) were removed from this correlation analysis to avoid the contamination from high DGL.
Thus the remaining dataset has small DGL component which is estimated by the equation (\ref{eq_DGL}).
Since the DGL subtracted spectra have two components (ZL and EBL),
the correlation between $SKY_i(\lambda )-DGL_i(\lambda )$ and $ZL_i(\lambda )$ from the equation (\ref{eq_ZL}) was investigated at each wavelength.
\begin{equation}SKY_i(\lambda )-DGL_i(\lambda ) = C(\lambda ) \cdot ZL_i (\lambda ) +  EBL(\lambda )  \label{eq_correlation}  \end{equation}
Two fitting parameters ($C(\lambda )$ and $EBL(\lambda )$) are obtained by the correlation, 
and the gradients of these correlations, $C(\lambda )$, give us a correction of our ZL estimation, and $y$-intercepts indicate the isotopic EBL.
The examples of correlations at some wavelengths are shown in Figure \ref{correlation}.
Best fit lines in Figure \ref{correlation} were obtained after 3$\sigma $ clipping for outlier rejection, and their gradients  $C(\lambda)$ are shown in Figure \ref{ZL_correction}.
Fairly good correlations can be seen and the gradient $C(\lambda)$ is basically consistent with unity within the error bars, indicating that the DIRBE ZL model is consistent with our data set.
However, the gradients at short wavelengths ($<$3.5 $\mu$m) are larger than unity by $\sim$5\%, indicating that the DIRBE ZL model underestimates the ZL intensity in these wavelengths.
This trend is consistent with the fact that the estimated ZL brightness by the DIRBE ZL model are smaller than other models by 5-8 \% at 1.25 $\mu$m and 11-14 \% at 2.2 $\mu$m \citep{Kelsall98}.

The dataset used in this study includes two types of errors, the statistical random error and the calibration error.
The statistic error includes the instrumental noise, uncertainty of the ZL estiation, and uncertainty of the DGL estimation.
As described in Paper I, the calibration error ($Err^{cal}(\lambda)$) is estimated to be 8\% at 1.8 $\mu$m, 16\% at 2.2 $\mu$m, and $<5$\% at $>2.5$ $\mu$m. 
The calibration error is larger than the statistical error at $<3.5$ $\mu$m.
Since propagations of these errors are different, these two errors were separately investigated in this correlation analysis.
First, the correlation analysis was conducted with only the statistical random errors, and EBL at each wavelength was obtained as $y$-intersects with errors, $EBL(\lambda ) \pm Err1(\lambda )$.
Errors owing to the calibration is estimated as $Err2(\lambda ) = EBL(\lambda ) \cdot Err^{cal}(\lambda )$,
and then these two types of errors were added to obtain the total error, $Err^{total}(\lambda ) =\sqrt{ Err1(\lambda )^2 + Err2(\lambda )^2}$.

\section{Result} \label{sec_result}
\begin{figure*}
  \begin{center}
    \FigureFile(160mm,100mm){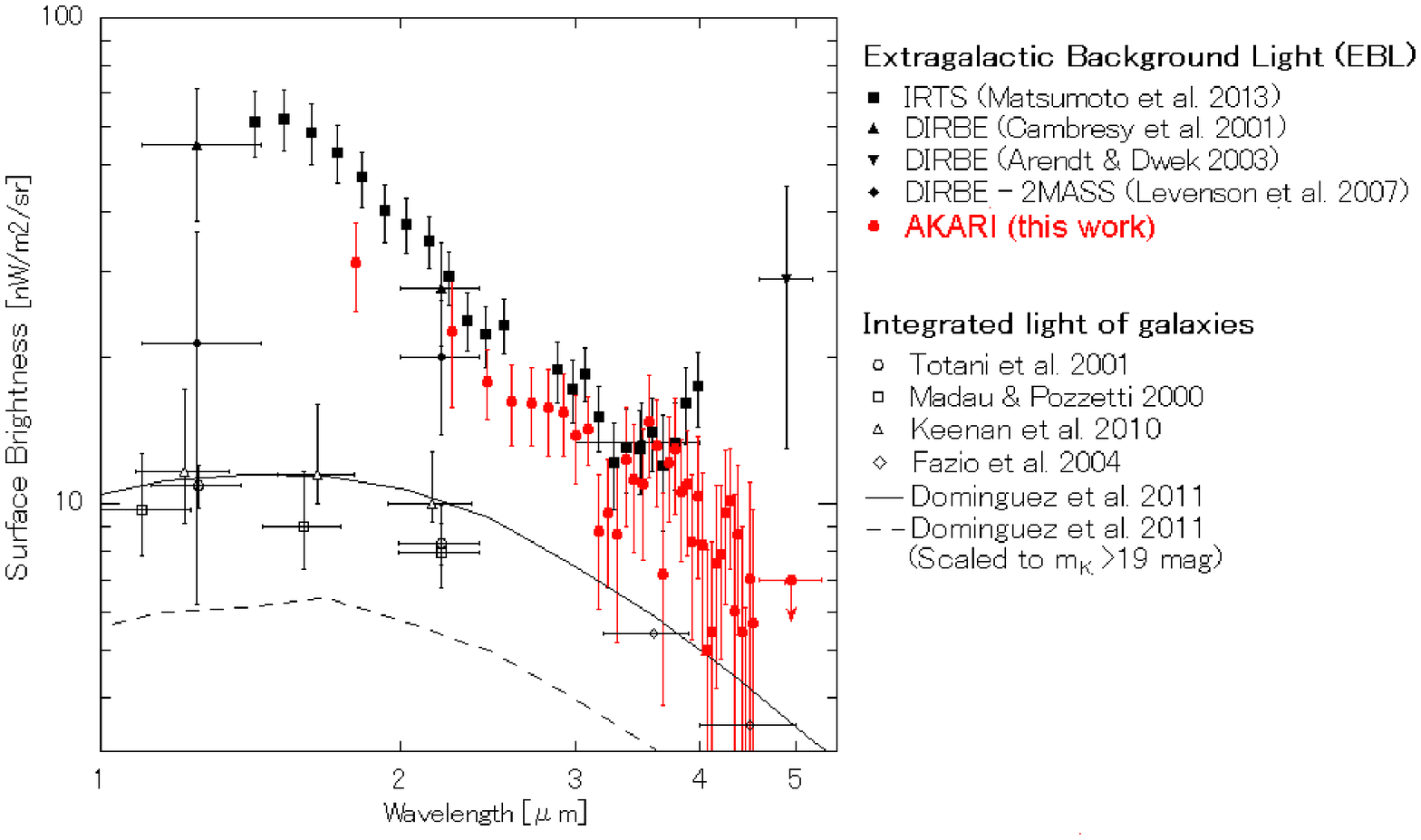}
  \end{center}
  \caption{Spectrum of EBL and integrated light of galaxies.
  Filled plots show EBL by various direct photometry from space including this study, and open plots shows the integrated light of galaxies by deep observations.
  Horizontal bars show the band widths of wide-band data.
  Solid curve shows a model spectrum of the integrated light of galaxies based on the observed evolution of the rest-frame K-band galaxy luminosity function up to redshift 4 \citep{Dominguez2011},
  and broken curve shows a scaled version of it in case of AKARI's detection limit of point sources ($m_K = 19$).  
  } 
  \label{CIB_all}
\end{figure*}

\begin{figure*}
  \begin{center}
    \FigureFile(160mm,100mm){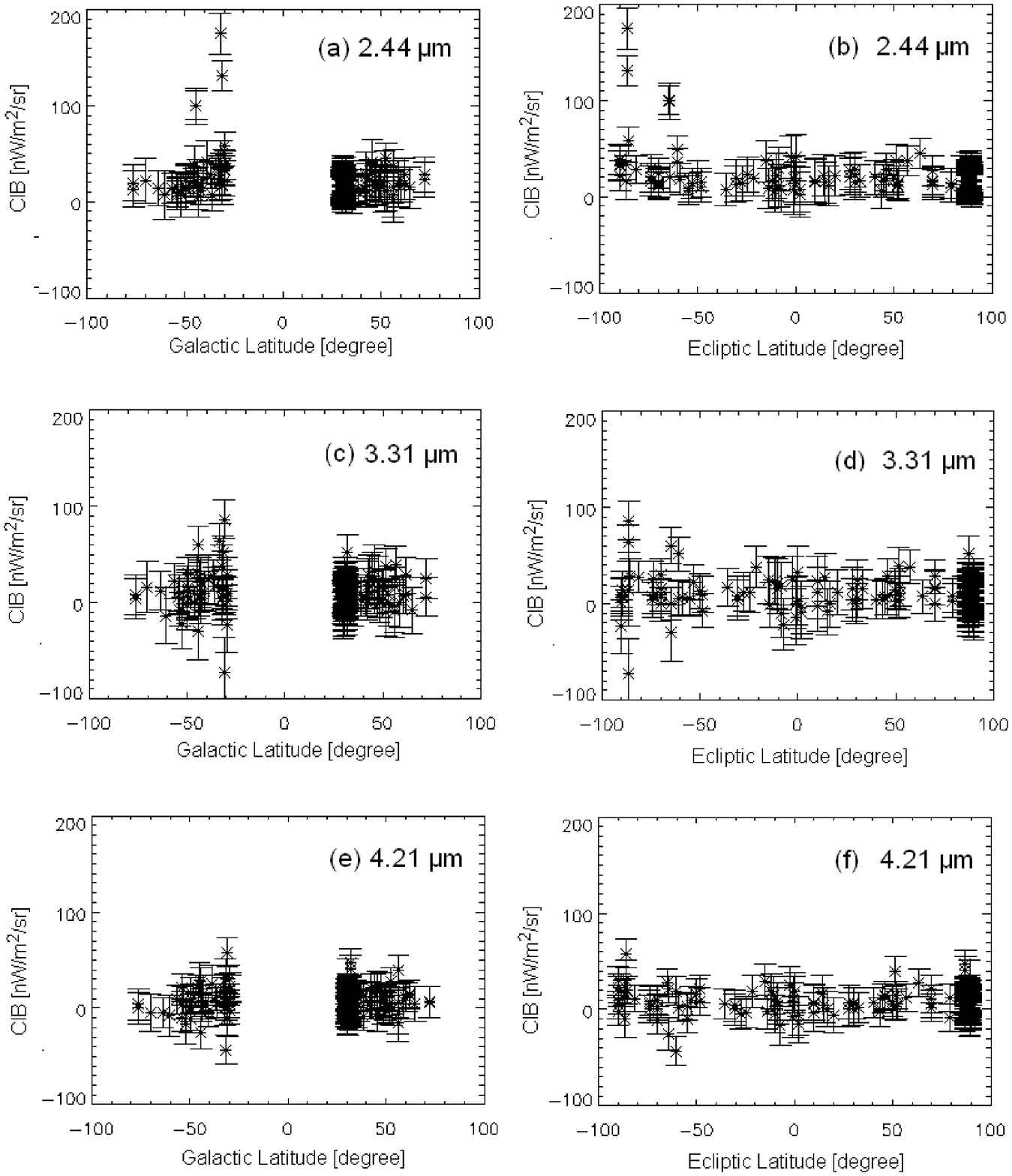}
  \end{center}
  \caption{Galactic latitude (a, c, e) and ecliptic latitude (b, d, f) dependence of EBL at 2.25 $\mu$m (a, b) , 3.31 $\mu$m (c, d), and  4.21 $\mu$m (e, f).
                  Error bars in this plot include only the statistic error ($Err1(\lambda )$). }
  \label{dependance}
\end{figure*}

Figure \ref{CIB_all} shows resultant EBL spectrum from our AKARI dataset with various previous results. 
Galactic and ecliptic latitude dependence of the obtained EBL was checked as shown in Figure \ref{dependance},
and the obtained EBL brightness is basically confirmed to be isotropic.
 
Our spectrum by AKARI/IRC clearly indicates the excess over the integrated light of galaxies 
and basically consistent with the revised IRTS result \citep{Matsumoto2013} and DIRBE results \citep{Cam01, Levenson07}.  
At the wavelength shorter than 3 $\mu$m, AKARI result shows a little lower brightness than the IRTS result. 
This difference could be attributed to the difference of the detection limits for the point source. 
At the K band, the limiting magnitude of AKARI dataset for removing point sources is 19 mag while that of IRTS is 10.5 mag. 
The difference of the integrated light of unresolved galaxies owing to this limiting magnitude difference is estimated to be $\sim$5 \nw\ \citep{Keenan10} which explains the difference of EBL brightness.

There is a gap in our result between 3.0 $\mu$m and 3.5 $\mu$m due to the difficulty of modeling the foregrounds in this wavelength region. 
As shown in Figure \ref{sky}, the diffuse sky brightness in this wavelength range include all three of foreground components 
(scattered sunlight of IPD, thermal emission from IPD, and DGL with 3.3 $\mu$m PAH band) with similar fractions, which makes the foreground modeling difficult. 
One possible explanation of the gap at around 3.3 $\mu$m is an overestimate of the 3.3 $\mu$m PAH band intensity. 
In fact, if we estimate DGL without the PAH band, the gap disappears and the EBL spectrum becomes smooth. 
This suggests that the correlation between PAH band emission and the 100 $\mu$m intensity at high Galactic latitude is fairly weaker than that at low Galactic latitude. 
As described in Paper II, the correlation between the PAH band and 100 $\mu$m intensity is confirmed only at $\mid b \mid <15^{\circ}$, 
and the DGL spectrum is estimated by extrapolation of this correlation to the higher Galactic latitude regions in our method. 
However, this assumption is obviously too simple. 
For example, UV radiation field at high Galactic latitude is weaker than that at Galactic plane \citep{Seon2011}, therefore the PAH molecules are less excited at high Galactic latitude than Galactic plane. 
This consideration indicates that the gap around 3.3$\mu$m could not be EBL origin but Galactic origin.

We obtained new spectral result of EBL at $>$4 $\mu$m, and 
we cannot confirm the excess over the integrated light of galaxies due to the large error bars. 
In addition, our result contradicts with the high EBL brightness at 4.9 $\mu$m by \citet{Arendt03}, 
but that data is highly uncertain since it is not an observed value but an estimated value from EBL at 1.25, 2.2, 3.5, and 100 $\mu$m.

\section{Discussion} \label{sec_discussion}
In Section \ref{sec_result}, we found NIR EBL observed with AKARI is fairly consistent with previous observations by COBE/DIRBE and IRTS/NIRS. 
How can we understand the excess of EBL from the integrated light of galaxies at $<$4 $\mu$m? 

At first we examine the possible origin in solar system. 
If there is an isotropic component in ZL, it cannot be subtracted by the correlation method in our study. 
One candidate of isotropic ZL component is a dust shell contingent on the Earth, but such a dense dust shell around the Earth must be detected already, if it exists. 
An isotropic diffuse background from the Oort cloud could be another candidate. 
However, the very blue spectrum toward 1 $\mu$m  cannot be generated by thermal emission from very cold dust ($<$30 K) at the Oort cloud. 
Scattered sunlight by the Oort cloud is also negligible because sunlight at $\sim10^4-10^5$ au is very weak.

The second possibility is Galactic origin. 
There may exist numerous faint stars in the Galactic halo which causes isotropic background. 
However, the negative detection of extended halo in external galaxies was reported by \citet{Uemizu1998}. 
Furthermore, the observed excess emission, $\sim$23 mag/arcsec$^2$ at H-band, can be easily detected for the external galaxies with HST/NICMOS \citep{Thompson2007a, Thompson2007b}, 
but no detection is reported yet. 
These considerations support that the observed excess emission is extragalactic origin.

Observation of TeV-$\gamma$ blazar is another problem for the extragalactic origin, 
since high level NIR EBL makes intergalactic space opaque for TeV-$\gamma$ photons \citep{Dwek05, Aha06, Aha07, Mazin07, Raue09}. 
However, recent discoveries of high redshift (z$>0.2$) TeV-$\gamma$ blazar  \citep{Ackermann2011} contradict with above standard scenario, and it requires a new physical process.
One idea is that cosmic rays produced by brazers can cross cosmological distances, and interact with NIR photons relatively close to the Earth, generating the secondary TeV $\gamma$-ray photons \citep{Essey2010}.
Another possible idea is that if TeV $\gamma$-ray photons are converted into Axion-like particles (ALP) and then regenerated to TeV $\gamma$-ray photons in our Galaxy.
In this case, TeV $\gamma$-ray photons should not suffer absorption effects while they propagate as ALPs \citep{Sanchez2009}.
If these processes work well, TeV $\gamma$-ray and EBL observations can coexist.

The spatial fluctuations of EBL were observed at 2.4, 3.2, and 4.1 $\mu$m by AKARI IRC imaging data \citep{Matsumoto2011} and at 3.6 and 4.5 $\mu$m by {\it Spitzer} IRAC imaging data \citep{Kashlinsky2005, Kashlinsky2007, Kashlinsky2012} in order to avoid uncertainty of ZL model, since ZL is known to be spatially smooth \citep{Pyo2012}. 
Observed fluctuations are consistent with each other and show significant large fluctuation at the angular scale larger than 100 arcsec, which cannot be explained by known foreground emission. 
The ratio of the EBL fluctuating power at $>100$ arcsec scale \citep{Matsumoto2011} to the absolute EBL spectrum (our result) can be obtained from the same instrument at the same season,
and we obtained $\delta I/I =$0.014 at 2.4 $\mu$m, 0.012 at 3.2 $\mu$m, and 0.0063 at 4.1 $\mu$m.
We also compare our EBL spectrum with the EBL fluctuating power at $>100$ arcsec scale obtained by {\it Spitzer} \citep{Kashlinsky2012}, obtaining $\delta I/I =$0.0064 at both 3.6 $\mu$m and 4.5 $\mu$m.
These results are consistent with $\delta I/I \sim$0.01 at any wavelengths from a theoretical estimation to predict the EBL fluctuations by the stracture formation at the early universe \citep{Fernandez2010},
supporting that both observed EBL spectrum excess and fluctuation have the same origin.

One of notable candidate of origin of the EBL excess and fluctuation is first stars of the universe \citep{San02, Sal03, Cooray04, Dwe05, Madau05, Mii2005, Fernandez06, Fernandez2010}. 
Our spectrum expands the previous IRTS and DIRBE spectra to 5.3 $\mu$m, and this blue EBL spectrum is consistent with stellar spectrum.
\citet{Dwe05} tried to explain the high EBL brightness at 4.9 $\mu$m \citep{Arendt03} by the redshifted H-$\alpha$ emission from the first stars at z$>$6, but this case is denied by our result.
However, recent detailed theoretical works \citep{Cooray2012a, Fernandez2012, Yue2013} indicate both of expected brightness and fluctuations are 10 times or more fainter than the observed ones, 
although spectral shape of excess emission and fluctuation is similar to theoretical prediction. 
\citet{Cooray2012b} suggested a model to explain the EBL fluctuation by the diffuse intrahalo light of galaxies, 
but it looks difficult to explain the EBL excess. 
The exotic origins related to the dark matter and/or the dark energy seems advantageous for the energetics point of view, but the theory on the origin of NIR EBL is not clear at present.

The origin of excess emission and fluctuation is still not clear, and new observations are highly expected to delineate their origin. 
The Cosmic Infrared Background ExpeRiment (CIBER) \citep{Bock2006, Zemcov2011} will provide the spectrum of the sky at 0.75-1.8 $\mu$m 
with the Low Resolution Spectrometer (LRS) \citep{Tsumura10, Tsumura2011} and fluctuation at 1.1 and 1.6 $\mu$m with the wide-field imagers \citep{Bock2012}
with a simultaneous observation of absolute brightness of ZL with the Narrow-Band Spectrometer (NBS) \citep{Korngut2013}. 
Observation from outside the zodiacal cloud is also highly required to conduct an ideal observation of EBL without the strong ZL foreground. 
A small infrared telescope, EXo-Zodiacal Infrared Telescope (EXZIT), has been proposed as one of instruments on a Solar Power Sail mission to Jupiter \citep{Matsuura02}. 
The measurement of NIR EBL at 5 au will be conducted in the 2020s.

\bigskip
This research is based on observations with AKARI, a JAXA project with the participation of ESA.
This research is also based on significant contributions of the IRC team.
We thank Mr. Arimatsu Ko (ISAS/JAXA) for discussion about the data reduction.
The authors acknowledge support from Japan Society for the Promotion of Science, KAKENHI (grant number 21111004 and 24111717).

\end{document}